%% file: bif-seqPhysicaD.tex
\documentclass{elsart}
\usepackage{amsmath}
\usepackage{amssymb}
\usepackage{epsfig}
\usepackage{changebar}
\usepackage{longtable}
\usepackage{amscd}
\usepackage{tabularx}

\begin{document}
\bibliographystyle{plain}



\input{bif-seq-titlepage}
                \def\thefootnote{\arabic{footnote}}
                        \setcounter{footnote}{0}


In their first edition of Fluid Mechanics \cite{llfluid1}, Landau and
Lifschitz proposed a route to turbulence in
fluid systems.  Since then, much work, in dynamical
systems, experimental fluid dynamics, and many other fields has been done concerning the
routes to turbulence.  In this paper, we present early results
from the first large statistical study of the route to chaos in a very
general class of high-dimensional, $C^r$, dynamical systems.  Our
results contain both some reassurances based on a wealth of previous
results and some surprises.  We conclude that, for high-dimensional discrete-time maps, the most probable route to chaos (in our general
construction) from a fixed point is via at least one Neimark-Sacker
bifurcation, followed by persistent zero Lyapunov exponents, and
finally a bifurcation into chaos.  We observe both the Ruelle-Takens scheme as
well as persistent $n$-tori, where $n \leq 2$ before the onset of chaos.
The point of this report is two-fold.  First, we propose a
manageable set of functions that will yield a better general
understanding of high-dimensional routes to chaos in a practical
sense.  Second we provide a survey of the routes to chaos in
this proposed set of mappings.  The goal is to select a common set of
mappings that exists someplace between real-world examples (the set we
propose can approximate a very general set of real world examples),
and the low-dimensional cases considered in rigorous mathematics.

Begin with an ordinary
differential equation in $R^k$ with a single real parameter $\mu$, $\frac{dv}{dt}
= F(\mu, U)$ where $F$ is as smooth as we wish and $U \subset R^k$ is compact.  At $\mu_0$ there exists a
fixed point, and at $\mu_c$, $\mu_0 < \mu_c$, $F$ is chaotic.  The
bifurcation sequence proposed by Landau \cite{llfluid1}
and Hopf \cite{hopfroutetochaos} is the
following: as $\mu$ is increased from $\mu_0$ there will exist a
bifurcation cascade of quasi-periodic solutions existing on higher and
higher dimensional tori until the onset of ``turbulence.''  In other
words, the solutions would be of the following type, $x_{\mu_1}(t) =
f(\omega_1, \omega_2)$, $x_{\mu_2}(t) =
f(\omega_1, \omega_2, \omega_3)$, ..., $x_{\mu_{k-1}}(t) = f(\omega_1,
\omega_2, \dots, \omega_k)$ for $\mu_i < \mu_{i+1}$, and where none of
the frequencies are rationally related.  However, Landau and Hopf's
notion of turbulence was high-dimensional, quasi-periodic flow.  Ruelle and Takens
\cite{randtturbulence} proposed both an alternative notion of
turbulence (the strange attractor) and an alternative route to turbulence in
a now famous paper.  Ruelle and Takens claimed that the Landau path was
highly unlikely from a topological prospective.  The basis for their claim originates in the work of
Peixoto \cite{peixotothrm} who has shown that quasi-periodic motion on $T^2$
(the $2$-torus) is non-generic\footnote{A property is generic if it
  exists on subset $E \subset B$, where $E$ contains a countable intersection of open sets that are
  dense in the original set $B$.} in the set of
$C^r$ vector fields.  However, Peixoto's theorem applies only to
flows on $T^2$ and not $T^k$ for $k>2$.  For diffeomorphisms of the circle,
irrational rotations make up the full Lebesgue measure set of
rotations.  Suspensions of such diffeomorphisms correspond to
quasi-periodic motion on $T^2$.  It would seem that
quasi-periodic motion of flows on $T^2$ would be high measure,
however, there is not a one-to-one correspondence between flows on
$T^2$ and discrete-time maps of the circle (e.g. the Reebs
foliation or \cite{palis_vf_approx}).  Further, quasi-periodic orbits of diffeomorphisms on the
circle are structurally unstable.  For flows on $T^2$, the structurally
stable, hyperbolic periodic orbits are topologically generic; however,
it is likely that quasi-periodic orbits are common in a measure
theoretic sense on $T^2$.  There remain many open questions regarding bifurcations
of periodic orbits, for example, a list of codimension $2$ bifurcations and their
status can be found at the end of section ($9.1$), page $397$ of \cite{kuzbook}.  How
this will all play out in practice is unclear and comprises a good
portion of motivation for our study.  Ruelle and Takens then went on to prove two results
for flows relevant to this report.  The first is a normal form theorem for the ``second'' Hopf
bifurcation for vector fields, or the ``first'' Hopf bifurcation for
maps (often referred to as the Neimark-Sacker bifurcation \cite{sacker}
\cite{naimark}).  This theorem gives a normal form analysis of the bifurcation of an invariant
circle of a flow, but it does not state the type of dynamic that will exist upon the loss of
stability of the invariant circle.  The second relevant result was that, given a
quasi-periodic solution $f(\omega_1, \dots, \omega_k)$ on $T^k$, $k
\geq 4$, in every $C^{k-1}$ small neighborhood of $(\omega_1, \dots,
\omega_k)$, there exists an open set of vector fields with a strange
attractor (prop. $9.2$ \cite{randtturbulence}).  These results were extended by
Newhouse, Ruelle and Takens \cite{nandrandtturbulence} who proved that a $C^2$
perturbation of a quasi-periodic flow on $T^3$ can produce strange
(axiom A) attractors, thus reducing the dimension to three for tori
with quasi-periodic solutions for which an open set of $C^2$ perturbations yield strange
attractors.  The basic scheme used by Takens, Ruelle and Newhouse was
to first prove a normal form theorem from periodic orbits to $2$-tori
in vector fields and then prove something about how the $2$-tori
behave under perturbations --- showing that bifurcations of $m$-tori
to $(m+1)$-tori will yield chaos since $3$-tori can be perturbed away to
axiom A chaotic attractors. 

If the story were only as simple as a disagreement between topological
and measure theoretic viewpoints, we would be in good shape.  However,
out of the complexity of the dynamics and the difficulties posed by
bifurcation theory regarding what happens to bifurcations of resonant
periodic orbits and quasi-periodic orbits, the field of quasi-periodic
bifurcation theory was born  \cite{broer_h_takens} \cite{braaksma1}.  We will not attempt to discuss the
history from where the above leaves off to the current state.  For those interested, see \cite{broerast},
\cite{chencinerandiooss1},  \cite{ioossconjecture},
\cite{eckmannpathtoturbulence}, \cite{broer_quasi_book}, \cite{iooss_los}, and \cite{iooss_adelmeyer}.  The question regarding the most
common route to turbulence is, in any but a very select set of specific
examples, still an open and poorly defined question.  Analytically piecing together
even what types of bifurcations exist en route to chaos has been slow
and difficult.  We contribute to the existing partial solution
in the following ways: we will provide a framework for numerical
analysis that does not have a priori tori built in; we will provide
evidence that the dominant scenario with respect to the cascade of bifurcations leading to chaos in many
high-dimensional dynamical systems will consist of various
bifurcations between periodic orbits of high-period and quasi-periodic
orbits.  Moreover, flip and fold bifurcations (due to real eigenvalues) occur with less and
less frequency as the dimension of the dynamical system is increased.

 

Let us begin by arguing why we believe that, as the dimension of
the dynamical system is increased, flip, fold,
any bifurcations due to real eigenvalues, and strong resonance bifurcations will be vanishingly rare
and the path to chaos from a fixed point in parameter space will consist of a soup of
various periodic orbits with high-period and quasi-periodic orbits.  Our intuition is born from results in the random
matrix theory of Girko \cite{girko1}, Edelman \cite{edelman1}, and Bai \cite{bai}.
The basic idea of their work is the following: given a square matrix whose
elements are real random variables drawn from a distribution with a
finite sixth moment, in the limit of infinite dimensions, the
normalized spectrum (or eigenvalues) of the matrix will converge to a uniform
distribution on the unit disk in the complex plane.  It is worth
noting that the convergence in measure is not absolutely continuous with respect to
Lebesgue measure.  Nevertheless, if the Jacobian of the map at the ``first'' bifurcation point
(i.e. the bifurcation from the fixed point) is a high-dimensional matrix
whose elements have a finite sixth moment, it is reasonable that the
bifurcation would be of type Neimark-Sacker (via a complex
eigenvalue) instead of a
flip or fold, with probability approaching unity as the dimension goes
to infinity.  Following this
line of thought, we, with W. Davis Dechert, provided strong numerical
evidence that the first bifurcation in typical high-dimensional
discrete-time dynamical systems would be that of the Neimark-Sacker
type \cite{albersroutetochaosI}.  Extending this analytically
for the $k^{th}$ bifurcation before the onset of chaos where $k>1$ is
made difficult (a proof for the $k=1$ case is not difficult  to construct) by the fact that the normal form and center manifold
theory for quasi-periodic bifurcation theory is a long way from
providing a general form(s) about which the results of
the random matrix theory could be applied (the codimension $2$
situation is not complete yet, and the codimension $3$ case is even
further from completion).  However, it is assumed that, in the end,
most bifurcations of periodic and quasi-periodic orbits can be captured
by some sort of Taylor series expansion (via a vector field
approximation or a suspension).  Though the linear term of the Taylor expansion will be degenerate and
the outcome of the bifurcation will be determined by contributions of higher-order terms, the degeneracies in the linear
term of the Taylor polynomial will nevertheless be of complex
eigenvalues --- leading to some sort of a bifurcation (yet to be
understood) from
quasi-periodic or periodic orbits to other quasi-periodic or periodic
orbits.   


A full statistical study is hampered by numerical accuracy issues
which may be fundamental problems associated with the existence of
neutral directions.  We will present the most clear picture of the
prototypical route to chaos in high dimensions for our set of
functions via a prototypical example.  For the understanding reached
in this report, we will employ bifurcation diagrams, phase plots, the
largest Lyapunov exponent and the full Lyapunov spectrum.  Numerical issues
related to the Lyapunov spectrum are the main problem with respect
to the statistical study, and we will discuss and display such
problems.  For information regarding why the numerical problems (such
as truly infinite convergence times) we
encounter may be unavoidable, see \cite{pughshubAMS} and
\cite{partialhyplese} for the latest with regard to the
existence of zero Lyapunov exponents and \cite{golub} \cite{highamnumaccuracy} with respect
to the linear algebra issues with computing them.
At the end of the report we will note the numerous extensions our
current line of study suggests.

The example we will present is typical (we will discuss what is not
typical about our example during the analysis) amongst the $500$ or so cases
(with this number of parameters and dimensions) we have observed in the sense that between the
first bifurcation and the onset of chaos, the only type of orbit that exists
is either quasi-periodic or periodic orbits with periods high enough such
that they are
indiscernible from quasi-periodic orbits.  It is worth noting that
the route to chaos we discuss in this report, which we believe is
the typical route in high dimensions, is considerably
different from what is observed at low dimensions.  For an intuitive
feel for the lower-dimensional cases, see \cite{albersroutetochaosI}.

A primary tool of analysis will be the Lyapunov exponent spectrum
(\cite{benn2} \cite{ruellehilbert}), since it is a good measure of the tangent space of
the mapping along its orbit.  Negative Lyapunov exponents correspond
to global stable manifolds or contracting directions; positive Lyapunov exponents correspond to global unstable manifolds or expanding directions and are (\cite{ruellehilbert}), in a computational
framework, the hallmark of chaos.  A zero exponent corresponds to a
neutral direction (although the story in this case is significantly
more complicated, see \cite{partialhyplese}); when the largest Lyapunov
exponent for a discrete-time map is zero and all other exponents are less
than zero, then there exists a neutral direction.  Lyapunov
exponents relate to quasi-periodic orbits and rotations in the
following way, neutral rotating directions correspond to a zero
Lyapunov exponent, and high-period orbits correspond to pairs of
negative Lyapunov exponents.  If the largest exponent is zero (while all others are negative),
then there exists a quasi-periodic orbit on the circle (on
$T^2$ in the flow).  If two exponents are zero, then there is a full
$2$-torus in discrete-time (a $3$-torus in the flow), and so on (for
specific examples of high-dimensional tori see \cite{broer_quasi_book}).  With respect to the
codimension $2$ bifurcations, there can also be as many as $10$ possibilities for zero Lyapunov exponents at bifurcation points
(again see page $397$ of \cite{kuzbook}), some of which correspond to real
and complex conjugate pairs of eigenvalues.  Besides the full Lyapunov
spectrum algorithm, we will also use an independent
version of a calculation of the largest Lyapunov exponent a la Wolf
et. al \cite{wolfle}.  Upon analysis of the example, it will be
obvious why we need such a calculation.  


The discrete-time mappings we will consider are single-layer, feed-forward neural networks of the form
\begin{equation}
\label{equation:net1}
 x_{t} = \beta_0 + \sum_{i=1}^{n}{{\beta}_i \tanh \left( s{\omega}_{i0} + s
 \sum_{j=1}^{d}{{\omega}_{ij} x_{t-j} } \right)} 
\end{equation}
which is a map from $R^{d}$ to $R$.  In eq.(\ref{equation:net1}), $n$ represents the number of hidden units or
neurons, $d$ is the input or embedding dimension of the system which
functions simply as the number of time lags, and $s$ is a scaling factor on the
weights.  The parameters are chosen in the following way: ${\beta}_i , {w}_{ij}
, x_{j} , s  \in {R}$,  where the $\beta_{i}$'s and ${w}_{ij}$'s are elements of weight matrices (which we
hold fixed for each case), $({x}_{0}, x_{1}, \ldots , x_{d})$
represent initial conditions, and $({x}_{t}, x_{t+1}, \ldots
, x_{t+d})$ represent the current state of
the system at time $t$.  We assume that the $\beta$'s are $iid$ uniform over $[0,1]$
and then re-scaled to satisfy $\sum_{i=1}^{n}{{{\beta}_{i}^{2}}} =
n$.  The ${w}_{ij}$'s are $iid$ normal with zero mean and unit variance.
The $s$ parameter is a real number and can be interpreted as the
standard deviation of the $w$ matrix of weights.  The initial $x_j$'s
are chosen $iid$ uniform on the interval $[-1,1]$.  

We would like to make a few notes with respect to the squashing
function, $\tanh()$.  For $x$ very
near $0$, the $\tanh(x)$ function is nearly linear.  Moreover, $\tanh(x)$, for
$|x| \gg 1$, will tend to behave much like a binary
function ($ \pm 1$).  Thus the scaling parameter $s$ will provide a
unique bifurcation parameter that will sweep from linear ranges to
highly non-linear ranges, to nearly binary ranges --- fixed points to
chaos to what seem like periodic cycles.  In this report however, we will cease consideration of our
networks long before the neurons become saturated.

Scalar neural networks like these are ``universal approximators,'' meaning they
can approximate many very general spaces of mappings (e.g. any $C^r$ mapping and its derivatives
to arbitrary accuracy, given enough neurons).  That scalar neural networks can approximate the mappings we
are interested in is a topic addressed in \cite{hor2}.  Combining the approximation theorems
of Hornik et al \cite{hor2} and the time-series embedding results of Takens
\cite{takensstit} shows the equivalence between our neural networks and the dynamical systems from compact sets in $R^n$ to
themselves (for specific arguments along these lines, see \cite{hypviolation}).

\begin{figure}
\begin{center}
\epsfig{file=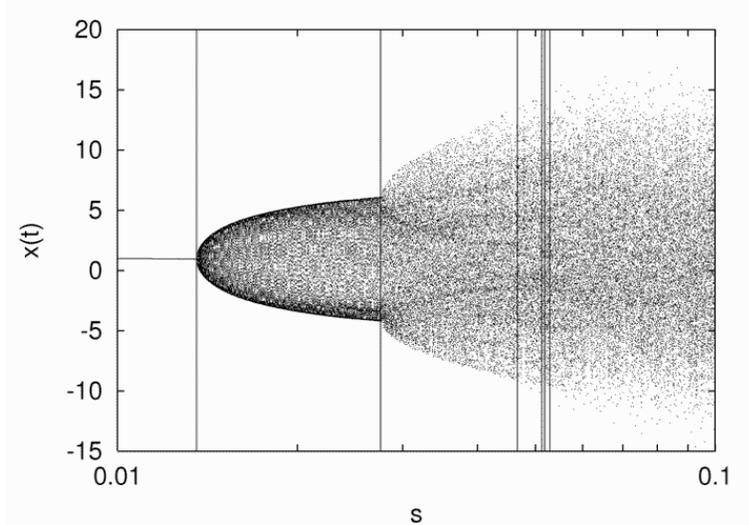, width=10.0cm, angle=0}
\caption{Bifurcation diagram for a typical network; $n=32$, $d=64$.}
\label{fig:bifdiagram}
\end{center}
\end{figure}


To circumvent the numerical stability issues that will be apparent shortly, we are required to
consider each network individually with four types of
figures: the largest Lyapunov exponent independently computed of the
full spectrum (it is considerably more
numerically stable than the algorithm for the full spectrum); a
standard bifurcation diagram; the full Lyapunov spectrum computed in
the standard manner; and phase-space diagrams.

Our choice of the number of neurons and the number of dimensions is
based on Fig. $(1)$ of \cite{albersroutetochaosI} and the compromise required by
computational time limits.  Considering Fig. $(1)$ of
\cite{albersroutetochaosI}, $32$ neurons puts our networks deeply in
the region of the set of neural networks that correspond to
highly complicated and chaotic dynamics.  The dimension, $64$ was
chosen because it was the highest dimension for which we could
reliably compute the nearly $500$ cases we surveyed.  The compute time
increases as a power of the dimension.  Thus as this time, we had
too few cases of $d=128$ and $d=256$ to make conclusive statements.    

Beginning with Fig. (\ref{fig:bifdiagram}), the standard bifurcation diagram, there are four important features to notice.  The first feature is the first
bifurcation, which occurs at $s=0.0135$ from a fixed point to some type
of limit cycle or torus.  A secondary bifurcation is clearly visible
at $s=0.02755$, the nature of this bifurcation is entirely a mystery
from the perspective of Fig. (\ref{fig:bifdiagram}).  Chaos seems to onset near
$s=0.05$, and has definitely onset by $s=0.06$, however the exact
location is difficult to discern.  Lastly, all of the dynamics between
the fixed point and chaos are some sort of $n$-torus ($n \geq 0$) type behavior.


\begin{figure}
\begin{center}
\epsfig{file=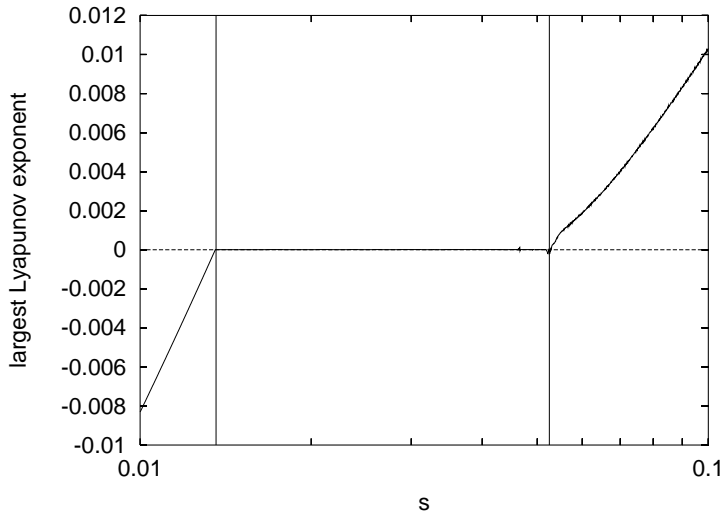, height=10cm, angle=270}
\caption{The Largest Lyapunov for a typical network; $n=32$, $d=64$.}
\label{fig:lle}
\end{center}
\end{figure}


Next, let us consider Fig. (\ref{fig:lle}) --- the largest Lyapunov
exponent versus variation in $s$ for the same case as Fig. (\ref{fig:bifdiagram}).  Again, as
expected, we see the first bifurcation at $s=0.0135$, in agreement
with the bifurcation diagram.  Figure (\ref{fig:lle}), however, gives
a clear picture of the onset of chaos, which occurs at $s=0.05284$.
The largest exponent is near zero between the first bifurcation and the onset of
chaos, providing evidence for the existence of at least one persistent complex
conjugate pair of eigenvalues with modulus one (assuming a Jacobian
can be constructed) --- e.g. a
persistent quasi-periodic orbit.  

\begin{figure}
\begin{center}
\epsfig{file=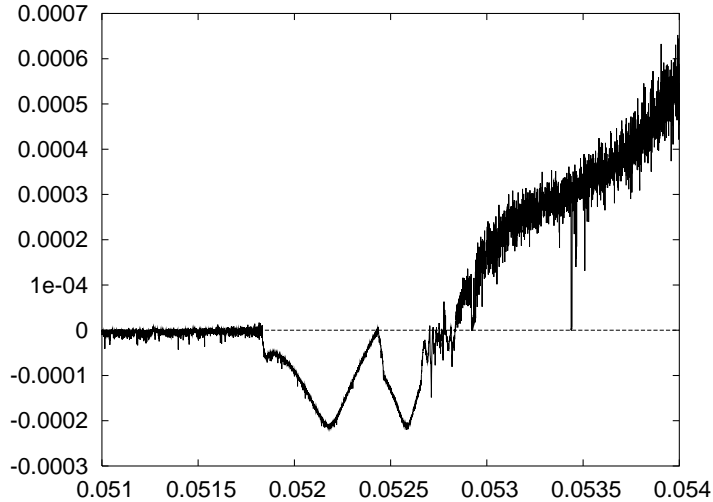, height=10cm, angle=270}
\caption{The Largest Lyapunov at the onset of chaos.}
\label{fig:onsetlle}
\end{center}
\end{figure}

Considering Fig. (\ref{fig:lle}), near the onset of chaos the exponent
becomes negative over a very short $s$ interval.  Ignoring all the intermediate bifurcations, let us briefly consider
the onset of chaos via Fig. (\ref{fig:onsetlle}).  Considering this
figure, there is an apparent periodic orbit, followed but what might
be a period doubling bifurcation, followed by what  could be a
complicated bifurcation structure.  Besides noting this for general
interest and completeness, we will refrain from a further discussion
of this small interval since this behavior seems to disappear for
high-dimensional networks and is not particularly related to the
point of this report.

\begin{figure}
\begin{center}
\epsfig{file=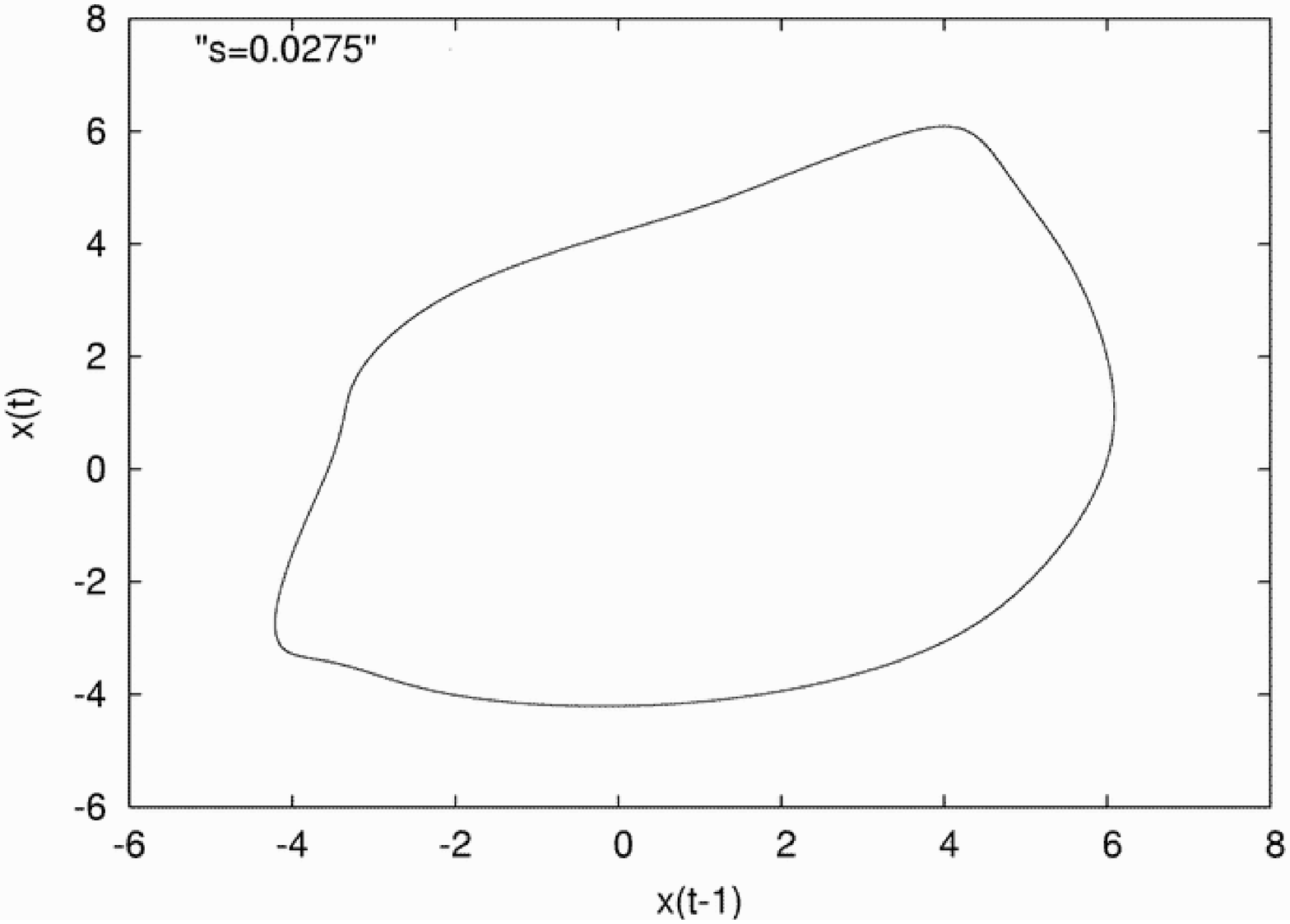, height=6.5cm, angle=270}
\epsfig{file=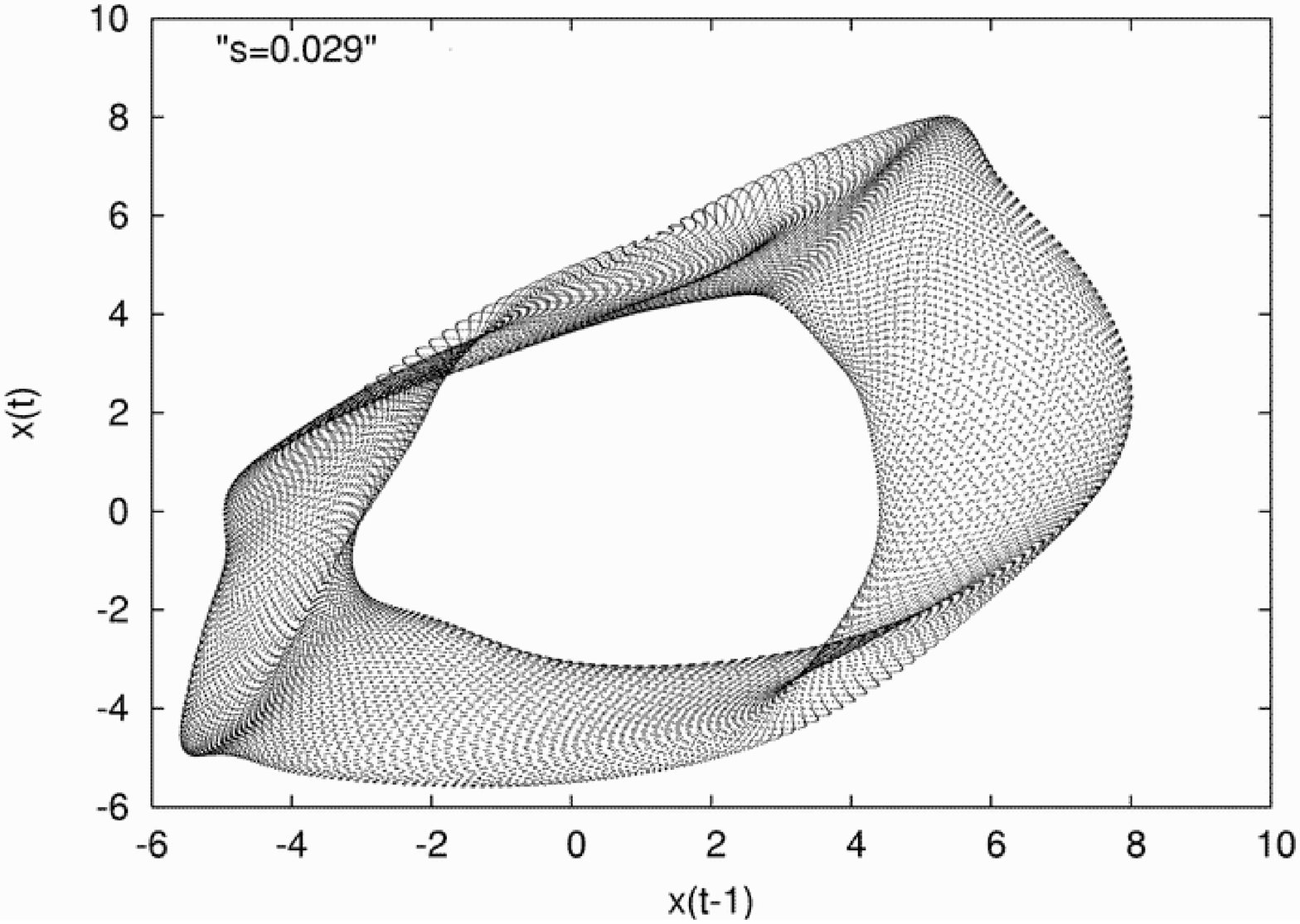, height=6.5cm, angle=270}
\caption{Phase plots on either side if the $2^{nd}$ bifurcation,
  $s=0.0275$ and $s=0.029$ respectively.  The bifurcation occurs at $s
  \sim 0.02754$.}
\label{fig:bif2}
\end{center}
\end{figure}

We will begin our presentation of phase-space figures at the
second bifurcation.  The second bifurcation is the obvious in
the bifurcation diagram, corresponding to the rapid change in the
attractor size near $s \sim 0.0275$.  The locations and nature
of these bifurcations can't be determined by a consideration of the
Lyapunov spectrum or the largest Lyapunov exponent, although
considering the Lyapunov spectrum it is clear that such a bifurcation
does occur.  The second bifurcation appears to be from a $1$-torus to a
$2$-torus as shown in Fig. (\ref{fig:bif2}).  

\begin{figure}
\begin{center}
\epsfig{file=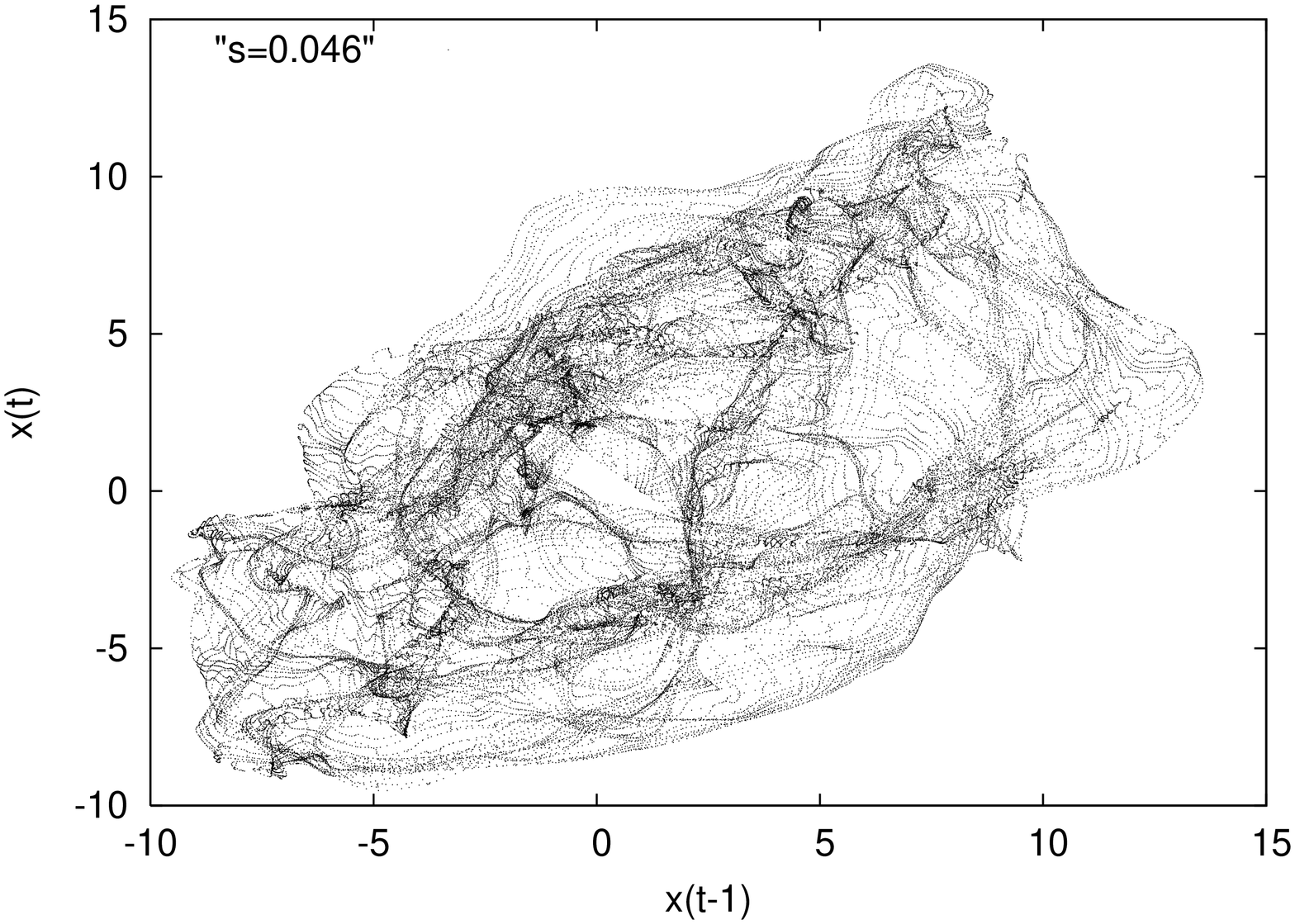, height=6.5cm, angle=270}
\epsfig{file=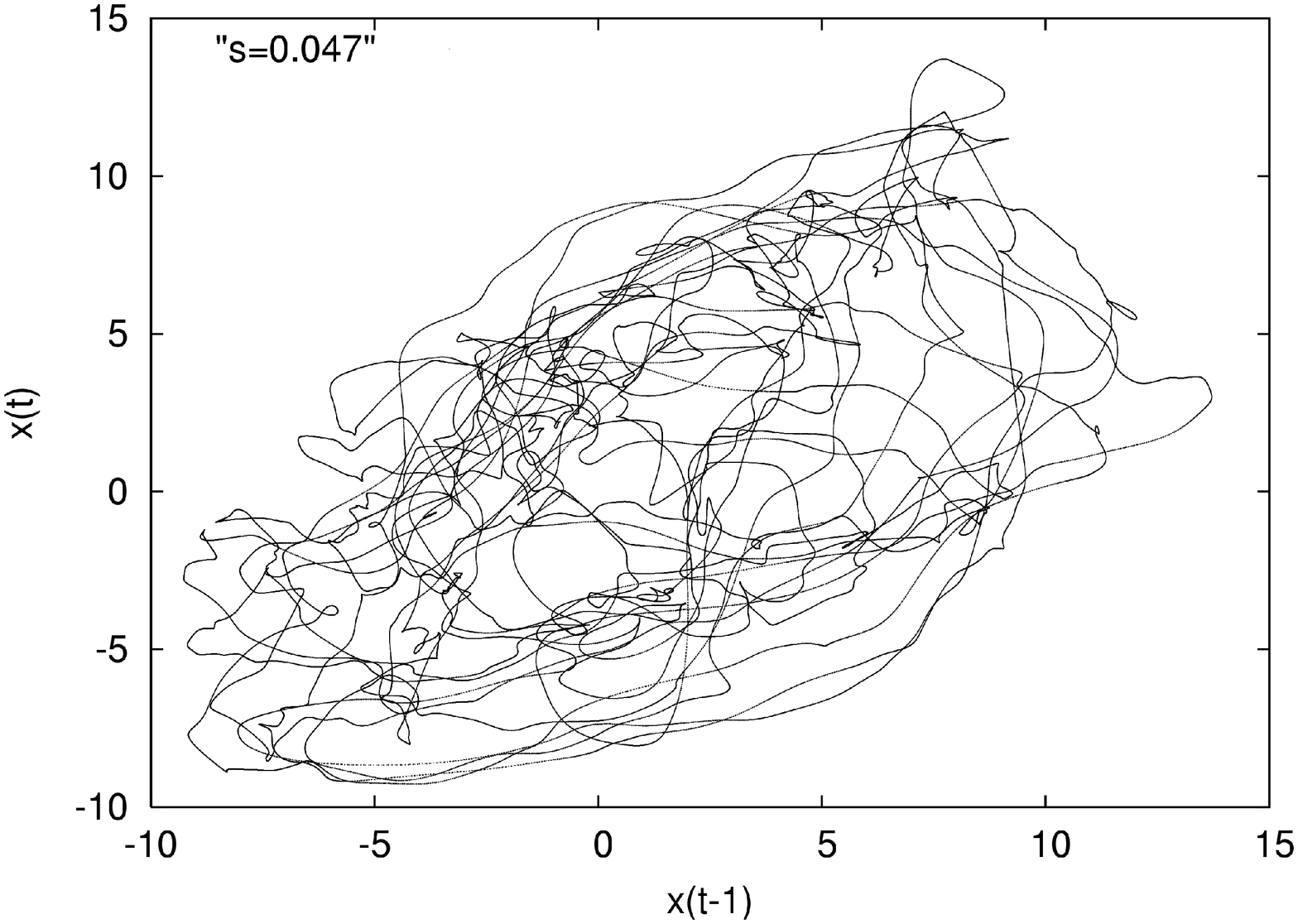, height=6.5cm, angle=270}
\caption{Phase plots on either side if the $3^{rd}$ bifurcation,
  $s=0.046$ and $s=0.047$ respectively.  The bifurcation occurs at
  $s \sim 0.04667$}
\label{fig:bif3}
\end{center}
\end{figure}

The phase-space plots on either side of the third bifurcation, which
is not clearly apparent in the bifurcation
diagram or in the Lyapunov spectrum as we will see, is depicted in
Fig. (\ref{fig:bif3}).  By this point, the smooth looking torus-like
object from Fig. (\ref{fig:bif2}) has become ``kinked''
significantly before the third bifurcation.  At the third bifurcation,
the torus-like, two-dimensional object becomes a
one-dimensional object.  Thus the third bifurcation is from a
$2$-torus to a $1$-torus and occurs at $s \sim 0.0466$.  The $1$-torus is a severely ``kinked''
quasi-periodic orbit.
  
\begin{figure}
\begin{center}
\epsfig{file=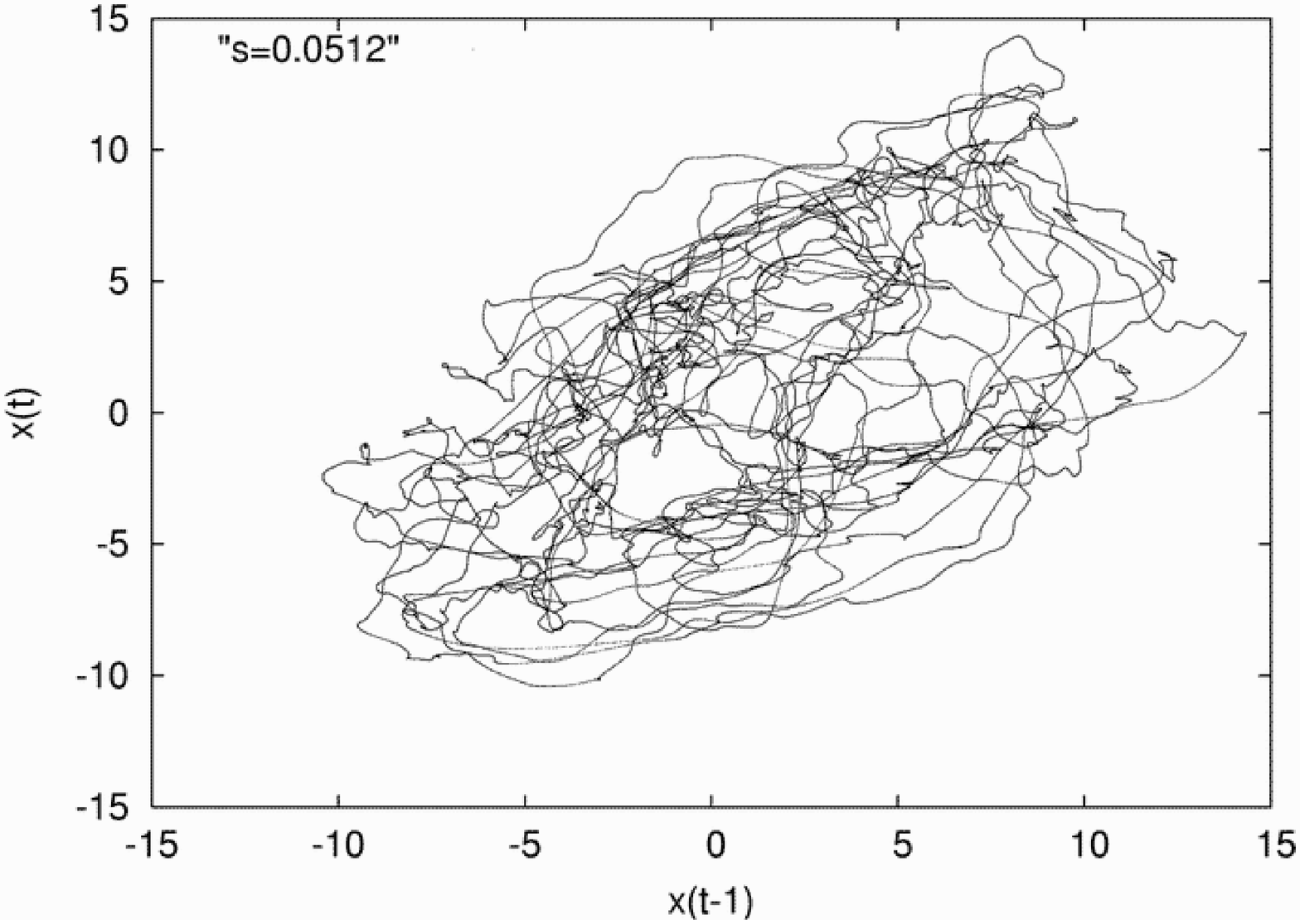, height=6.5cm, angle=270}
\epsfig{file=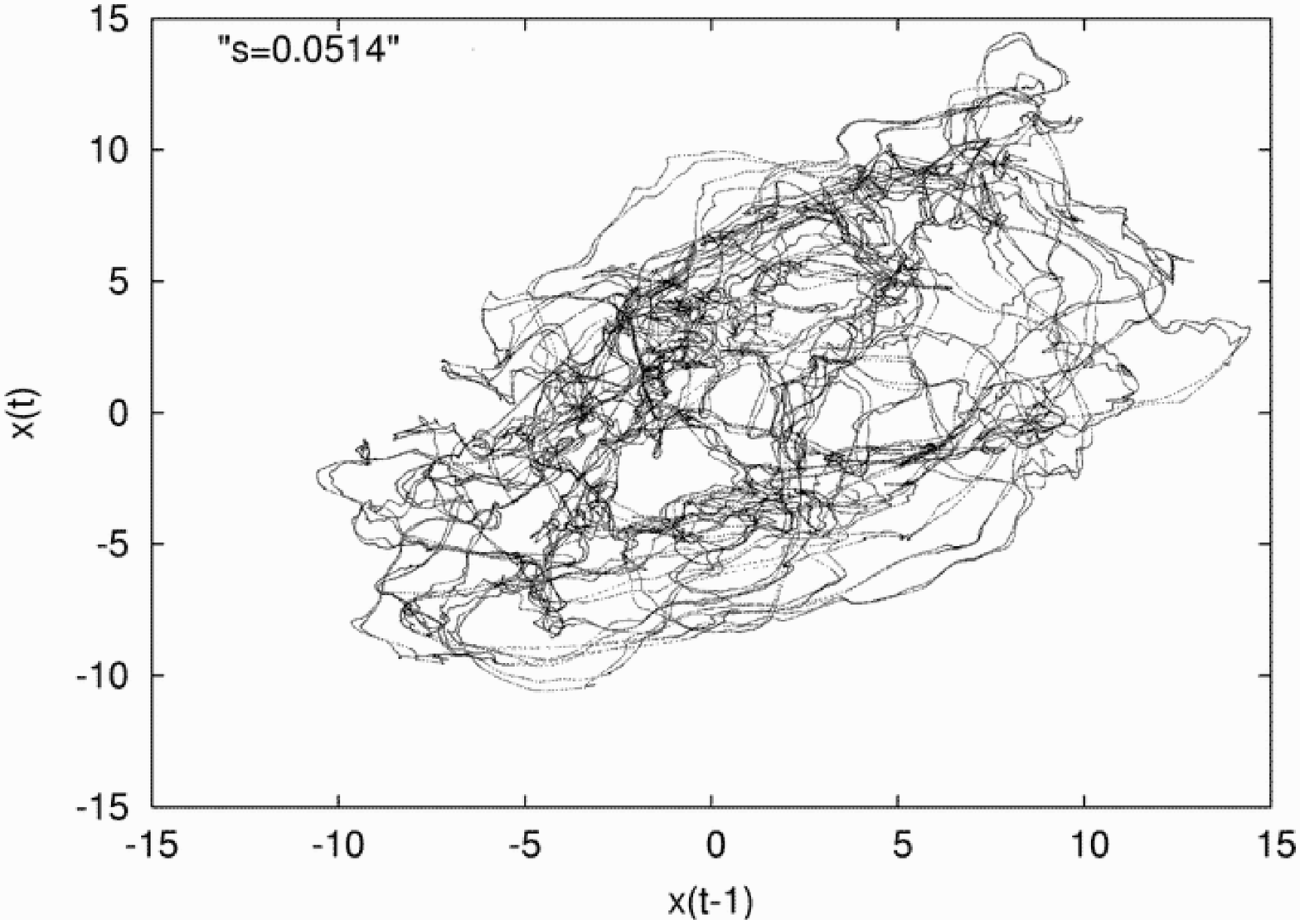, height=6.5cm, angle=270}
\caption{Phase plots on either side of the $4^{th}$ bifurcation,
  $s=0.0512$ and $s=0.0514$ respectively.  The bifurcation occurs at
  $s \sim 0.05124$}
\label{fig:bif4}
\end{center}
\end{figure}

Considering the phase-space plots on either side of the fourth
bifurcation in Fig. (\ref{fig:bif4}), one might conclude that there
our example has undergone a period doubling of the quasi-periodic
orbit.  An analytical explanation of such a bifurcation is yet an
open problem, but it is likely a ``Neimark-Sacker-Flip'' bifurcation.
We will refrain from a further discussion of this bifurcation,
directing the interested reader to chapter $9$ of \cite{kuzbook}.

\begin{figure}
\begin{center}
\epsfig{file=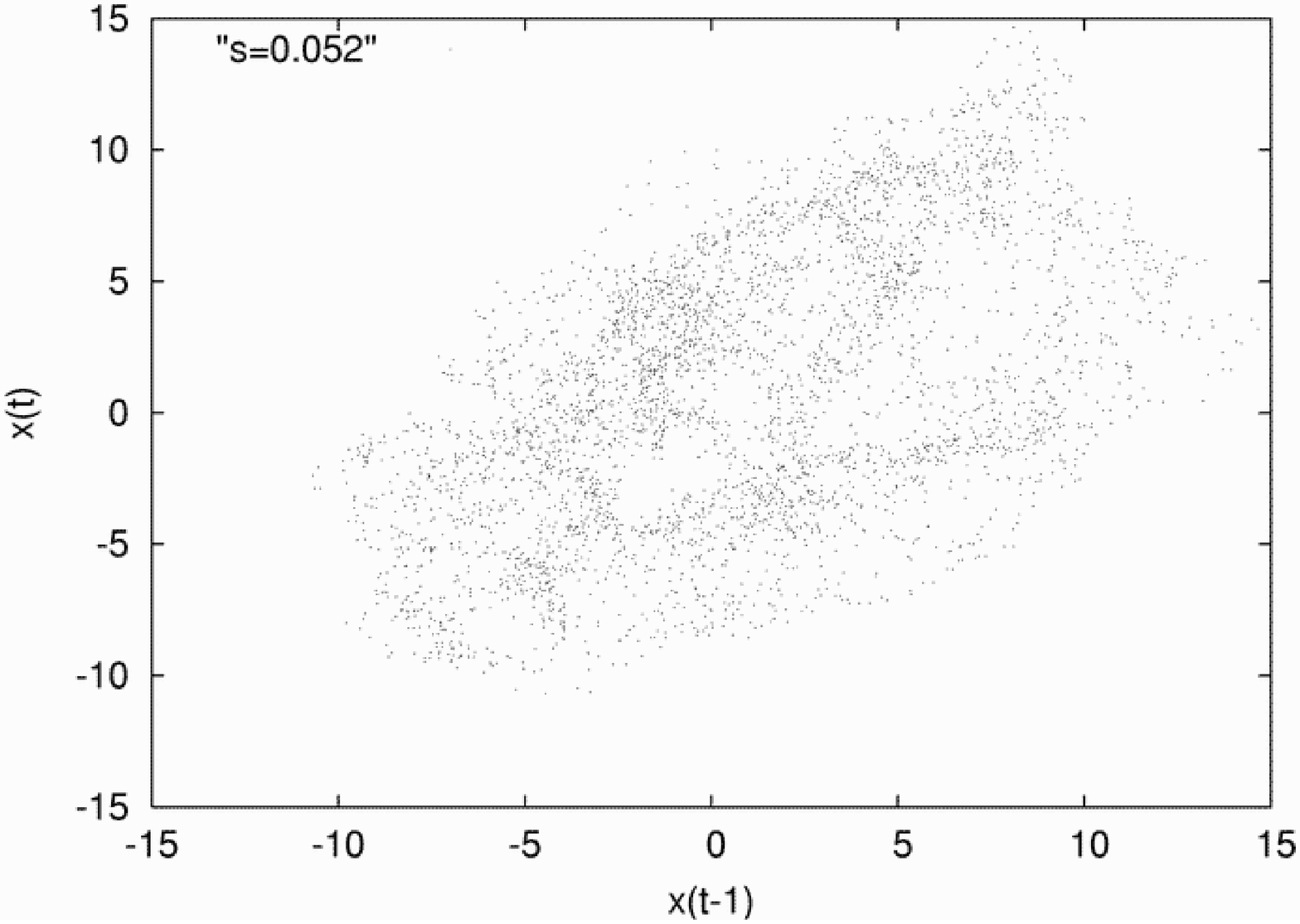, height=6.5cm, angle=270}
\epsfig{file=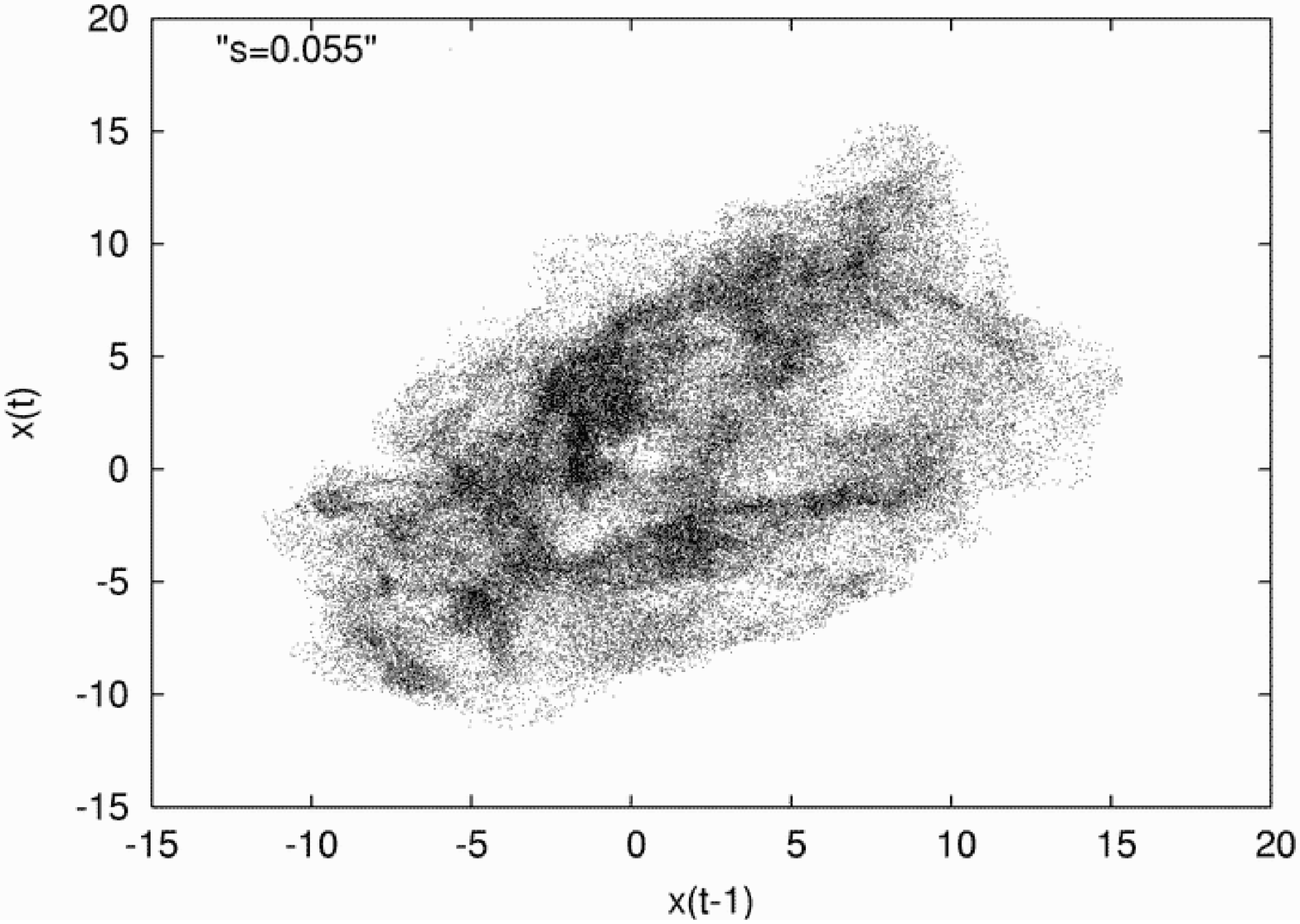, height=6.5cm, angle=270}
\caption{Phase space plot near the $6^{th}$ bifurcation --- $s=0.052$
  and $s=0.053$ respectively.  The bifurcation occurs at $s \sim 0.05294$.}
\label{fig:bif6}
\end{center}
\end{figure}

We will refrain from showing figures for the fifth bifurcation, simply
noting that it is a bifurcation from this quasi-periodic orbit on the
$1$-torus to a high-period, periodic orbit.  Rather, we will skip to
what we will call the sixth bifurcation.  Figure (\ref{fig:bif6})
demonstrates the final bifurcation into chaos from a high-period,
periodic orbit.  However, in our particular example, considering
Fig. (\ref{fig:onsetlle}), just before the onset of chaos, there is a
likely a sequence of bifurcations.  We will not belabor this
further, as little insight is gained from a further consideration, and
this sequence of bifurcations just before the onset of chaos appears
to be increasingly rare as the dimension increases.

\begin{figure}
\begin{center}
\epsfig{file=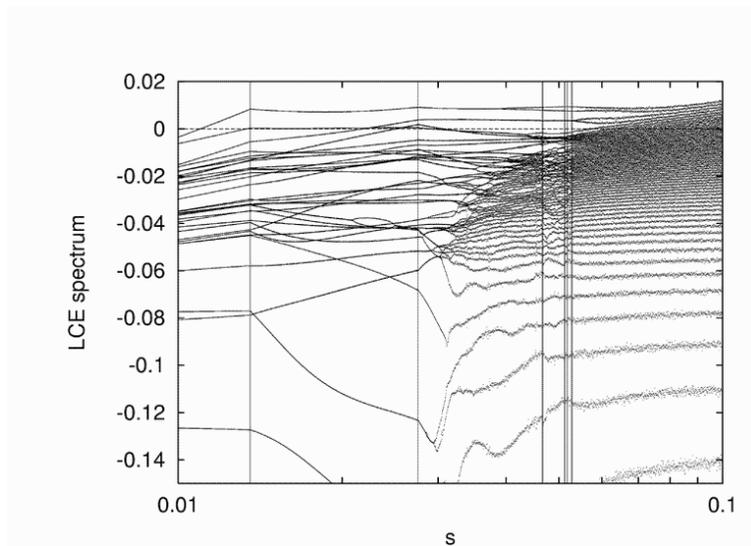, height=10cm, angle=270}
\caption{Lyapunov spectrum the (typical) network; $n=32$, $d=64$.}
\label{fig:lce}
\end{center}
\end{figure}


Before discussing what can be gleaned from the  Lyapunov spectrum data, we
must make a few comments with respect to the data plotted in Fig. (\ref{fig:lce}).  If one were to pick an $s$ value along the $s$
interval $(0, 0.03)$ and count the number of Lyapunov
exponents, it is likely that one could often only identify roughly half the
number of exponents that might be expected.  There is a simple reason
for this, if one were to consider the eigenvalues of the variational
differential equation,\footnote{See page 36 of \cite{pesinlebook} for a
definition of the variational differential equation.  In practical
terms, we are referring to something akin to the eigenvalues of the
local ``Jacobian'' of the network at a specific time.} nearly all the
eigenvalues are complex conjugate pairs, and hence
the pairs of nearly equal Lyapunov exponents.  This in a sense,
implies a lot of local rotation, but as this can depend upon the
coordinate system, little can be said in this regard.  Nevertheless,
from a practical point of view, this implies sink behavior in all
directions.  The point
is that the ``largest'' exponent is actually pair of nearly equal exponents.
This is true of at least the ten largest ``exponents,'' and often this is
true of all of the ``exponents'' on $(0, 0.03)$.  The Neimark-Sacker style
bifurcations to periodic orbits and tori are due to those pairs of
exponents.  If the bifurcations were of
real eigenvalues, there would not be pairs of Lyapunov exponents of
equal magnitude.

On the $s$ interval $\sim (0, 0.03)$, nearly all the exponents come in
pairs, all the lines represent two exponents.  Much of this is related
to the numerical stability issues related to the linear algebra
engines used to calculate the exponents.  This is a fundamental and
common problem in matrices with many nearly equal, and hence
degenerate eigenvalues.  This issue lies in the standard algorithm's
ability to accurately orthogonalize the eigenvalues and eigenvectors,
for more information see \cite{golub} and \cite{highamnumaccuracy}.  At $s \sim
0.03$ the negative exponents begin to split apart.  This cascade
continues to occur from the most negative exponents to the largest
exponents, in that order, up to the onset of chaos.  By the onset of
chaos, there exists $64$ distinct exponents, one for each dimension.  In fact, by
$s \sim 0.052$, all the exponent pairs are observably split.  It is possible
that this is the
first time this phenomena has been observed.  There are several interpretations
for such behavior; we will sketch our interpretation.  In the interval
where the exponents exist in nearby pairs, the dynamics are dominated by
rotation in nearly all eigen-directions, and the many nearly equal local
eigenvalues cause numerical stability problems.

As the $s$ parameter is increased, the,
rotating directions become less degenerate, and the lower
exponents begin to separate.  This phenomena has very little to do
with the overall dynamics when the
eigenvalues are strongly negative since these
directions are so strongly contracting that they are not observed
on the attractor after transients die out.  As the $s$ parameter is further increased,
more pairs of Lyapunov exponents separate until the neutral directions
begin to separate --- this is the break up of the tori and the onset
of turbulence.\footnote{With the break up of the commensurate
  directions, our numerical stability returns, see \cite{albersnil} or \cite{hypviolation}
  for more regarding the dynamics after this transition}  We are
displaying an example of the Ruelle-Takens scenario which contrasts
the Landau and Hopf turbulence model.  Strange attractors are not a collection of interacting
quasi-periodic orbits or a rotating soup, but rather, distinct directions of expansion,
contraction \textit{a la} axiom A, and a little bit of rotation (neutral directions).  This
example, and Fig. (\ref{fig:lce}) display this distinction in a very nice
way.  This cascade from all commensurate to all incommensurate
Lyapunov exponents was unexpected.  

The first bifurcation is relatively
easy to identify in Fig. (\ref{fig:lce}), it occurs at a relative maximum of the largest
overlaying set of exponents in the spectrum at $s=0.0135$.  This signals the
Neimark-Sacker type first bifurcation, as expected.  In our experience, the first
bifurcation can be determined from the full spectrum reliably.  The
second bifurcation is also, relatively obvious at $s=0.02755$, and
corresponds to the
relative maximum of the six (three pairs) largest exponents.  This would signal the
existence of a torus of some type, but due to the numerical stability
issues it is difficult to make a definitive statement regarding the dynamics.
Nevertheless, by the second bifurcation point, there clearly exists at least a $2$-torus.  Near the third bifurcation the Lyapunov spectrum becomes
significantly more complicated --- due to the numerical accuracy
issues we dare not make a claim with respect to how many exponents are
actually zero; however, it appears as if there are several.  The onset of chaos
is vague, but chaos can clearly be identified by $s=0.8$.
Likewise, considering Figs. (\ref{fig:lce}) and (\ref{fig:lle}), the
respective largest Lyapunov exponents begin to be
indistinguishable at $s=0.08$.  The secondary bifurcations cannot be
identified, however, and it is this problem that has hindered the
full statistical study. 


The previous information can be summarized in the following table for
which the location of the various bifurcations is given as observed
via the bifurcation diagram, the largest Lyapunov exponent, the phase
space diagrams, and the Lyapunov spectrum:

\begin{tabularx}{\linewidth}{|X|X|X|X|X|X|X|}
\hline
Bif. number & $s$ at the $1^{st}$ & $s$ at the $2^{nd}$ & $s$
at the $3^{rd}$ & $s$ at the $4^{th}$ & $s$ at the $5^{th}$  & $s$ at the $6^{th}$ \\
\hline
Largest Lyapunov exponent & $0.0135$ & --- & --- & --- & 0.05183 & $0.05289$ \\
\hline
Bif. diagram & $0.0135$  & $0.02755$ & --- & --- & --- & $0.0525 $  \\
\hline
Phase space diagram & $0.0135$ & $0.02754$ & $0.04667$ & $0.05124$ &
$0.05183$  & $0.05289$ \\
\hline
Lyapunov spectrum &  $0.0135$ &  $0.02755$ & --- & --- & --- & $0.053$ \\
\hline 
Transition to: & $T^1$ & $T^2$ & $T^1$ & $T^1$ & $T^0$ & Chaos \\

\hline
\end{tabularx}
 
\vspace{0.5cm}

Putting this all together into a unified picture, it is likely that the first bifurcation, of
type Neimark-Sacker, occurs at $s=0.0135$.  The second bifurcation
occurs at $s = 0.02754$ is a bifurcation from $T^1$ to $T^2$.  The
third bifurcation, which occurs at $s=0.04667$ is a bifurcation from
$T^2$ back to $T^1$.  The fourth bifurcation appears to be a
period-doubling like bifurcation from $T^1$ to $T^1$.  The fifth
bifurcation is a bifurcation from the quasi-periodic orbit to a very
high-period orbit.  Following this is a very subtle sequence
of bifurcations followed by the onset  of chaos at $s=0.05289$.

All the other observed cases with $d=64$ were variations on a theme.
We have observed more and fewer bifurcations between the first bifurcation and chaos, but all the
bifurcations are relatively similar.  Period-doubling cascades and such
routes are rare --- they occur in less than one percent of
$64$-dimensional networks with our weight distribution (i.e. we did
not  observe a period doubling route to chaos in the $500+$ networks
we considered for this report).  Only five percent of the first
bifurcations are due to real eigenvalues, and there never exist cascades
of multiple real bifurcations.  The most common route to chaos we
observed is a cascade of bifurcations between $T^1$, $T^2$, and
high-period, periodic orbits; $3$-tori are rare in our experience, we
never observed one in nearly $500$ cases.  


One strength of our methods and results is, unlike topologically based
results, our neural network
framework allows not only a practical means of analyzing topological
results, but it contains, in a manageable way, the supposed pathological examples.  Due to the
mappings that neural networks can approximate, if the spectrum of
Lyapunov exponents of a $d$-dimensional network is contained in
$[\chi_1,  \chi_d]$, then likely
there exists at least one path through parameter space such that any network can
be transformed into a $T^d$ torus with all Lyapunov exponents being
zero (we haven't proved such a result, however).  If one were to
stratify the networks by their spectra, the aforementioned torus would
be but a point along the interval $[\chi_1,  \chi_d]$, and thus, in
this sense, unusual.  There are other stratifications of the networks
that can be made, such as the map $\phi:R^{n(d+2)} \rightarrow
\Sigma(\tanh())$, where $\Sigma(\tanh())$ is the set of neural
networks with $\tanh()$ as the squashing function.  The point is, we
are presenting a practical framework that yields numerical
observations regarding common routes to chaos.  For example, the number of
constraints required to achieve a $d$-torus might be high\footnote{The
number of constraints for a $d$-torus might also be low, as requiring
area preservation, or a local Jacobian to have an average determinant equal to
$\pm 1$ is, in a sense, a singe constraint.}, and the $T^d$ torus would likely exist on a surface in parameter space
of much lower dimension than the ambient parameter space.  However,
since the neural networks can approximate $m$-tori for $m \leq d$, it
is possible to study the transitions, and the likelihood of such
transitions, and the persistence of high-dimensional tori in a
practical way.  From our
experience perturbing away $2$-tori and limit cycles requires drastic
parameter variation, however, we never observe $3$-tori consistent
with the prediction of Newhouse, Ruelle and Takens.


As this paper is intended as a survey of our space of neural networks
along the route to chaos, let us list current and future directions of
work:
\begin{itemize}

\item[i.] a specific numerical analysis of each of the bifurcations in
  this example that can be compared to the theory; specifically
  the second and third bifurcations which are likely of low codimension;

\item[ii.] an analytical normal form calculation --- again starting
  with the early bifurcations along the route;

\item[iii.] a study of Lyapunov spectrum calculation technique a la \cite{deelepaper}, \cite{lcecomp_methods}, \cite{qrlce},
  \cite{benn2}, or \cite{shimnag} --- these networks form a nice set of
  high-dimensional mappings to study Lyapunov spectrum
  calculation schemes since our mappings are high dimensional, not pathological, and are not, as high-dimensional
  maps go, computationally intensive to use; very often, upon the
  presentation of a new Lyapunov spectrum computation algorithm the
  test cases are very low dimensional dynamical systems for which
  numerical stability is rarely a problem;

\item[iv.] a brute force --- by hand --- statistical study of the
  bifurcation sequences in these networks;

\item[v.] a more numerically accurate Lyapunov spectrum
  calculation routine that can be used for a full statistical analysis
  of the routes to chaos in this set of dynamical systems;

\item[vi.] a systematic investigation of the dependence on the
  number of neurons;

\item[vi.] a systematic investigation of the sensitivity of the weight
  distribution;

\item[vii.] constraints on weight distributions which lead to dynamics
similar to a particular physical phenomenon, leading to a better
understanding for how special a physical phenomena is relative to a
general function space.

\end{itemize}

Our results help extend the current analytical results in
the sense that we have a practical way of observing transitions
along an interval in parameter space.  Further, we present a framework such
that the parameter set for which the claimed non-generic tori and
persistent quasi-periodic behavior can be more concretely understood.
Based on our observations, in high dimensions, the quasi-periodic route
to chaos, often with a cascade of bifurcations is
the dominant route.  We observe what is predicted to occur according
to the Ruelle-Takens route, but the bifurcation cascade is
significantly more complicated than $T^0 \rightarrow T^1 \rightarrow
T^2 \rightarrow \mathit{chaos}$.  The bifurcation cascade often
involves successive bifurcations between  tori with dimension$\leq 2$.

We would like to
thank R. A. Bayliss and W. D. Dechert for many helpful discussions and advice.  D. J. Albers
would like to give special thanks to J. P. Crutchfield for guidance,
many fruitful discussions, wonderful insight and feedback, and support.  The computing for the project was done
on the Beowulf cluster at the Santa Fe Institute and was partially
supported at the Santa Fe Institute under the Networks, Dynamics
Program funded by the Intel Corporation under the Computation,
Dynamics, and Inference Program via SFI's core grant from the National
Science and MacArthur Foundations.  Direct support for D. J. Albers
was provided by NSF grants DMR-9820816 and PHY-9910217 and DARPA
Agreement F30602-00-2-0583.

\bibliography{dstexts,partialhyperbolicity,lyapunovexponents,neuralnetworks,nilpotency,topology,analysis,structuralstability,computation,me,physics,bifurcationtheory,probability}


\end{document}

%% file: bif-seq-titlepage.tex
\begin{frontmatter}



\title{Routes to chaos in high-dimensional dynamical systems: A
  qualitative numerical study}


\author[uwdop,sfi]{D. J. Albers}
\ead{albers@santafe.edu}
\ead[url]{http://www.santafe.edu/$\sim$albers}
\author[uwdop]{J. C. Sprott}
\ead{sprott@physics.wisc.edu}
\ead[url]{http://sprott.physics.wisc.edu/sprott.htm}


\address[uwdop]{Department of Physics, University of Wisconsin, 
  Madison; 1150 University Avenue, Madison, WI 53706-1390}
 
\address[sfi]{Santa Fe Institute; 1399 Hyde Park Road, Santa Fe, NM 87501}

\begin{abstract}
This paper examines the most probable route to chaos in high-dimensional dynamical systems in a very general
computational setting.  The most probable route to chaos
in high-dimensional, discrete-time maps is observed to be a sequence
of Neimark-Sacker bifurcations into chaos.  A means for determining and understanding the degree to which the Landau-Hopf
route to turbulence is non-generic in the space of $C^r$ mappings is outlined.
The results comment on previous results of Newhouse, Ruelle, Takens,
Broer, Chenciner, and Iooss.  
\end{abstract}

\begin{keyword}


PACS:
05.45.-a  
05.45.Tp  
89.75.-k  
89.20.Ff 

\end{keyword}

\end{frontmatter}